\documentclass[12pt]{article}
\usepackage{graphicx}
\def \app{D_{\pi \pi}}

\def \bea{\begin{eqnarray}}
\def \beq{\begin{equation}}

\def \cn{Collaboration}

\def \eea{\end{eqnarray}}
\def \eeq{\end{equation}}
\def \ite{{\it et al.}}

\def \s{\sqrt{2}}

\def \3half{\frac{3}{2}}

\begin{document}

\begin{flushright}
EFI 03-19 \\
hep-ph/0305144 \\
May 2003 \\
\end{flushright}
  
\renewcommand{\thesection}{\Roman{section}}
\renewcommand{\thetable}{\Roman{table}}
\centerline{\bf A Comment on ``Weak Phase $\gamma$ using Isospin Analysis and} 
\centerline{\bf Time-dependent Asymmetry in $B^0 \to K_S\pi^+\pi^-$''
\footnote{To be published as a Comment in Physical Review Letters}}
\medskip
\centerline{Michael Gronau\footnote{Permanent Address: Physics Department,
Technion -- Israel Institute of Technology, 32000 Haifa, Israel.}}
\centerline{\it Enrico Fermi Institute and Department of Physics}
\centerline{\it University of Chicago, Chicago, Illinois 60637}
\bigskip

\begin{quote}
We comment on a recent suggestion for measuring $\gamma$ in $B^0(t)\to
K_S\pi^+\pi^-$. A difficulty is pointed out in relating 
electroweak penguin and tree amplitudes in $B^+\to K^0\pi^+\pi^0$, 
which is crucial for an implementation of this method.
We clarify the necessary condition for such a relation. 
\end{quote}

\leftline{\qquad PACS codes:  12.15.Hh, 12.15.Ji, 13.25.Hw, 14.40.Nd}

\bigskip
In a recent interesting Letter \cite{DSS} Nilendra Deshpande, Nita Sinha 
and Rahul Sinha propose to use 
$B\to K\pi\pi$ decays in order to determine the weak phase $\gamma$. 
Their method is based on the CP asymmetry in $B^0(t)\to K_S\pi^+\pi^-$, 
and on an isospin triangle relation among the three 
amplitudes for $B^+ \to K^0(\pi^+\pi^0)_e,~B^0\to K^0(\pi^+\pi^-)_e$ and 
$B^0\to K^0(\pi^0\pi^0)_e$, in which the two pions are in an even angular 
momentum state. A crucial assumption of this method is that
electroweak penguin and tree amplitudes contributing to 
$B^+ \to K^0(\pi^+\pi^0)_e$ involve a common strong phase.
Such a property was shown to hold in the SU(3) symmetry limit
for the $I=3/2$ amplitude in $B\to K\pi$ \cite{NR,GPY}, and in the isospin 
symmetry limit for the $I=2~B\to\pi\pi$ amplitude \cite{GPY,BF}.  

The purpose of this short comment is to clarify the general condition 
under which tree and electroweak amplitudes in charmless $B$ decays can 
be related to each other. We will show that this condition
is not fulfilled in the case studied in \cite{DSS}. 

The effective Hamiltonian describing charmless $\Delta S=1$ (or $\Delta 
S=0$) decays \cite{BBL} consists of current-current operators $Q_1$ and $Q_2$,
QCD penguin operators $Q_i,~i=3-6$, and electroweak penguin (EWP) operators 
$Q_i,~i=7-10$. The operators $Q_1$ and $Q_2$, multiplying Wilson coefficients 
$c_1$ and $c_2$, respectively, and CKM coefficients 
$V^*_{ub}V_{us}$ (or $V^*_{ub}V_{ud}$), will be named tree operators. 
EWP operators involve CKM factors $V^*_{tb}V_{ts}$ 
(or $V^*_{tb}V_{td}$). The EWP operators $Q_9$ and $Q_{10}$ with the 
dominant Wilson coefficients, $c_9$ and $c_{10}$, have the same 
(V--A)(V--A) structure as the tree operators, and would have approximately 
the same matrix elements if they had also identical flavor SU(3) and 
isospin structure. 

In order to find out when such a relation holds, one decomposes all four 
quark operators of the form $(\bar bq_1)(\bar q_2 q_3)$ transforming as 
${\overline{\bf 3}}\otimes {\bf 3}\otimes {\overline{\bf 3}}$ in terms of 
a sum of ${\overline {\bf 15}}$, ${\bf 6}$ and ${\overline {\bf 3}}$
\cite{Zepp}. The representation ${\overline{\bf 3}}$ appears both 
symmetric (${\overline{\bf 3}}^{(s)}$), and antisymmetric 
(${\overline{\bf 3}}^{(a)}$) under the interchange of $q_1$ and $q_3$.
The tree and electroweak parts of the $\Delta S=1$ Hamiltonian
are \cite{GPY}:
\beq\label{T}
{\cal H}_T = -\frac{G_F}{\s}V^*_{ub}V_{us}
\left [\frac{c_1-c_2}{2}(\overline{\bf 3}^{(a)}_0 + {\bf 6}_1) +
\frac{c_1+c_2}{2}(\overline {\bf 15}_1 + \frac{1}{\sqrt2}\overline 
{\bf 15}_0 - \frac{1}{\sqrt2}\overline{\bf 3}^{(s)}_0)\right ]~~,
\eeq
\beq\label{EWP}
{\cal H}_{\rm EWP} = -\frac{G_F}{\s}\frac{3V^*_{tb}V_{ts}}{2}\left [
\frac{c_9-c_{10}}{2}(\frac{1}{3}\overline{\bf 3}^{(a)}_0 + 
{\bf 6}_1) + \frac{c_9+c_{10}}{2}(-\overline {\bf 15}_1 
-\frac{1}{\sqrt2}\overline {\bf 15}_0
-\frac{1}{3\sqrt2}\overline{\bf 3}^{(s)}_0 )\right ]~~,
\eeq
where subscripts denote the isospin of corresponding operators. Note that
both the ${\bf 6}$ and $\overline {\bf 15}$ operators include a $\Delta I=1$
component. 

Eqs.~(\ref{T}) and (\ref{EWP}) imply proportionality relations between EWP 
and tree operators transforming as $\overline {\bf 15}$ and ${\bf 6}$   
\cite{MG}: 
\bea\label{15}
{\cal H}_{\rm EWP}(\overline{\bf 15}) &=& 
-\frac32 \frac{c_9+c_{10}}{c_1+c_2} 
\frac{V^*_{tb}V_{ts}}{V^*_{ub}V_{us}} 
{\cal H}_T(\overline{\bf 15})~,\\
\label{6}
{\cal H}_{\rm EWP}({\bf 6}) &=& \frac32 \frac{c_9-c_{10}}{c_1-c_2}
\frac{V^*_{tb}V_{ts}}{V^*_{ub}V_{us}}{\cal H}_T({\bf 6})~.
\eea
The two proportionality constants are approximately renormalization 
scale independent, and are approximately equal in magnitudes but differ 
in sign \cite{BBL},
\beq
\frac{c_9+c_{10}}{c_1+c_2} \approx \frac{c_9-c_{10}}{c_1-c_2}~~.
\eeq
Therefore, EWP and tree amplitudes in $B$ decay processes which obtain 
contributions from either the $\overline{\bf 15}$ or the 
${\bf 6}$ operator, but not from both, are proportional to each other 
and involve a common strong phase. This property does not hold when the
two operators contribute because of the opposite signs in Eqs.~(\ref{15}) 
and (\ref{6}). In this case EWP and tree amplitudes involve the sum and
difference of two complex amplitudes and have different strong phases.

In the case of $B\to (K\pi)_{I=3/2}$ \cite{NR,GPY}, the $K$ and $\pi$ are
in an S-wave state, which is symmetric under an interchange of the two SU(3) 
octets. This state is a pure ${\bf 27}$. The only SU(3) operator 
which contributes to this transition is the $\overline{\bf 15}$. 
Consequently, the EWP and tree amplitudes 
are proportional to each other in the SU(3) approximation. The same holds 
true in the isospin symmetry limit for the EWP and tree amplitudes of 
$B\to (\pi\pi)_{I=2}$, since only the $\overline{\bf 15}$ contains a 
$\Delta I=3/2$ component \cite{GPY,BF}.
On the other hand, in $B^+ \to K^0(\pi^+\pi^0)_e$ studied in \cite{DSS}
the final state has $I=3/2,~S=1$ and can be in a ${\bf 27}$ and 
in a ${\bf 10}$, to which the $\Delta I =1$ components of both
the $\overline{\bf 15}$ and the ${\bf 6}$ operators contribute. Hence, 
the condition for proportional EWP and tree amplitudes and for a common 
strong phase does not hold. Although this proportionality does not follow 
from symmetry considerations alone, it would be interesting to study 
possible dynamical assumptions which can lead to such a situation. 

\medskip
I thank Jonathan Rosner for useful discussions, and I am grateful to
the Enrico Fermi Institute at the University of Chicago for its kind 
hospitality. This work was supported in part by the United States 
Department of Energy through Grant No.\ DE FG02 90ER40560.

\def \ajp#1#2#3{Am.\ J. Phys.\ {\bf#1}, #2 (#3)}
\def \apny#1#2#3{Ann.\ Phys.\ (N.Y.) {\bf#1}, #2 (#3)}
\def \app#1#2#3{Acta Phys.\ Polonica {\bf#1}, #2 (#3)}
\def \arnps#1#2#3{Ann.\ Rev.\ Nucl.\ Part.\ Sci.\ {\bf#1}, #2 (#3)}
\def \art{and references therein}
\def \cmts#1#2#3{Comments on Nucl.\ Part.\ Phys.\ {\bf#1}, #2 (#3)}
\def \cn{Collaboration}
\def \cp89{{\it CP Violation,} edited by C. Jarlskog (World Scientific,
Singapore, 1989)}
\def \econf#1#2#3{Electronic Conference Proceedings {\bf#1}, #2 (#3)}
\def \efi{Enrico Fermi Institute Report No.}
\def \epjc#1#2#3{Eur.\ Phys.\ J.\ C {\bf#1}, #2 (#3)}
\def \ib{{\it ibid.}~}
\def \ibj#1#2#3{~{\bf#1}, #2 (#3)}
\def \ijmpa#1#2#3{Int.\ J.\ Mod.\ Phys.\ A {\bf#1}, #2 (#3)}
\def \ite{{\it et al.}}
\def \jhep#1#2#3{JHEP {\bf#1}, #2 (#3)}
\def \jpb#1#2#3{J.\ Phys.\ B {\bf#1}, #2 (#3)}
\def \mpla#1#2#3{Mod.\ Phys.\ Lett.\ A {\bf#1} (#3) #2}
\def \nat#1#2#3{Nature {\bf#1}, #2 (#3)}
\def \nc#1#2#3{Nuovo Cim.\ {\bf#1}, #2 (#3)}
\def \nima#1#2#3{Nucl.\ Instr.\ Meth.\ A {\bf#1}, #2 (#3)}
\def \npb#1#2#3{Nucl.\ Phys.\ B~{\bf#1}, #2 (#3)}
\def \npps#1#2#3{Nucl.\ Phys.\ Proc.\ Suppl.\ {\bf#1}, #2 (#3)}
\def \PDG{Particle Data Group, K. Hagiwara \ite, \prd{66}{010001}{2002}}
\def \pisma#1#2#3#4{Pis'ma Zh.\ Eksp.\ Teor.\ Fiz.\ {\bf#1}, #2 (#3) [JETP
Lett.\ {\bf#1}, #4 (#3)]}
\def \pl#1#2#3{Phys.\ Lett.\ {\bf#1}, #2 (#3)}
\def \pla#1#2#3{Phys.\ Lett.\ A {\bf#1}, #2 (#3)}
\def \plb#1#2#3{Phys.\ Lett.\ B {\bf#1}, #2 (#3)}
\def \prl#1#2#3{Phys.\ Rev.\ Lett.\ {\bf#1}, #2 (#3)}
\def \prd#1#2#3{Phys.\ Rev.\ D\ {\bf#1}, #2 (#3)}
\def \prp#1#2#3{Phys.\ Rep.\ {\bf#1}, #2 (#3)}
\def \ptp#1#2#3{Prog.\ Theor.\ Phys.\ {\bf#1}, #2 (#3)}
\def \rmp#1#2#3{Rev.\ Mod.\ Phys.\ {\bf#1}, #2 (#3)}
\def \yaf#1#2#3#4{Yad.\ Fiz.\ {\bf#1}, #2 (#3) [Sov.\ J.\ Nucl.\ Phys.\
{\bf #1}, #4 (#3)]}
\def \zhetf#1#2#3#4#5#6{Zh.\ Eksp.\ Teor.\ Fiz.\ {\bf #1}, #2 (#3) [Sov.\
Phys.\ - JETP {\bf #4}, #5 (#6)]}
\def \zpc#1#2#3{Zeit.\ Phys.\ C {\bf#1}, #2 (#3)}
\def \zpd#1#2#3{Zeit.\ Phys.\ D {\bf#1}, #2 (#3)}

\end{document}